\documentclass[aps,pra,epsfigure,showpacs,twocolumn,superscriptaddress,longbibliography,floatfix]{revtex4-1}
\usepackage{dcolumn}    
\usepackage{bm} 
\usepackage{graphicx}
\usepackage{amsmath}    
\usepackage{latexsym}
\usepackage{amsfonts}   
\usepackage{amssymb}
\usepackage{array}      
\usepackage{epsfig}
\usepackage{txfonts}
\usepackage{color}
\usepackage[colorlinks=true,linkcolor=blue,urlcolor=blue,citecolor=blue]{hyperref}
\usepackage{hyperref}
\usepackage{epstopdf}
\usepackage{dsfont}

\begin{document}
 \title{Spin quantum heat engines with shortcuts to adiabaticity}

\author{Bar\i\c{s} \c{C}akmak}
\email{cakmakb@gmail.com}
\affiliation{Department of Physics, Ko\c{c} University, \.{I}stanbul, Sar\i yer 34450, Turkey}
\affiliation{College of Engineering and Natural Sciences, Bah\c{c}e\c{s}ehir University, Be\c{s}ikta\c{s}, Istanbul 34353, Turkey}
\author{\"{O}zg\"{u}r E. M\"{u}stecapl{\i}o\u{g}lu}
\email{omustecap@ku.edu.tr}
\affiliation{Department of Physics, Ko\c{c} University, \.{I}stanbul, Sar\i yer 34450, Turkey}

\begin{abstract}
We consider a finite-time quantum Otto cycle with single and two-spin-$1/2$ systems as its working medium. In order to mimic adiabatic dynamics at a finite time, we employ a shortcut-to-adiabaticity technique and evaluate the performance of the engine including the cost of the shortcut. We compare our results with the true adiabatic and non-adiabatic performances of the same cycle. Our findings indicate that the use of the shortcut-to-adiabaticity scheme significantly enhances the performance of the quantum Otto engine as compared to its adiabatic and non-adiabatic counterparts for different figures of merit. 
\end{abstract}

\pacs{}
\date{\today}

\maketitle
\section{Introduction}
Thermodynamics is a field of study exploring the relationship between various forms of energy and energy transfer. A practical motivation behind it was the development of thermal engines in XIXth century. Thermodynamical laws are found to be significant for computational machines as well, where the interplay of information, heat and work has been well-established~\cite{landauer_irreversibility_1961}. While these developments are for large systems, there is a strong interest to extend them for nano and atomic technologies. Classical thermodynamics has been extended on the scale of nanometers~\cite{chamberlin_big_2014}. For systems in the quantum domain, systematic studies have been started more recently and led to the rapidly advancing field of quantum thermodynamics~\cite{JPA_Goold,ContPhys_Anders,arXiv_Alicki}. In contrast to their classical counterparts however, quantum heat engines (QHEs) have marginal power outputs and hence they are currently far from being a practical stimulus for quantum thermodynamics. A crucial bottleneck in the operation of QHEs is the slowness of quantum analogs of expansion and compression processes. Here we explore how some shortcuts for such processes can be used in a QHE and if the engine can still be efficient when the energetic costs of the shortcuts are taken into account.

Despite their limited practical value, QHEs are pivotal to reveal fundamental limits on the operation
of quantum machines from a thermodynamical perspective. Accordingly, many proposals to realize them can be found in the literature~\cite{PRE_Quan1,PRE_Quan2,Entropy_Kosloff,JCP_Kosloff,PRL_Kieu,EPJD_Kieu,PRE_Ferdi,PRE_Ferdi2,EPJP_Selcuk,PRA_Gabriele,Entropy_Deffner,PRE_Thomas,PRE_Lee,PRL_Obinna,PRA_Delgado,PRE_Gong2}, some
of which take into account finite-time engine cycles~\cite{EPJD_Selcuk,PRE_Zheng,QST_Marco,PRE_MustecapliogluExp,feldmann_quantum_2006}. 
In addition, the effects of the profound quantum
nature of the QHE such as such as cooperativity~\cite{NJP_Jaramillo,arXiv_Klimovsky,NJP_Wolfgang,NatComm_Campisi,PRE_Mojde}, coherence and correlations~\cite{PRE_Wolfgang,PRE_Dorfman,arXiv_Guff} on the performance of QHEs have also been investigated. Following the demonstration
of a single-ion engine cycle~\cite{Science_Obinna}, genuine QHE experiments have been reported~\cite{arXiv_Paterson} with a single spin in a nuclear magnetic resonance (NMR) set up~\cite{arXiv_Paterson,de_assis_quantum_2018}, a cold Rb atom~\cite{zou_quantum_2017}, and an nitrogen-vacancy (NV) center in diamond~\cite{klatzow_experimental_2017}.
Here we consider finite-time QHEs with single and two-spin working systems subject to shortcuts to adiabaticity.

The techniques of shortcuts to adiabaticity (STA) deal with the problem of making an adiabatic transformation of a quantum system at a finite-time by external manipulations on the system, which otherwise needs to be made infinitely slow. There are different strategies to achieve this goal, and each of them has its own advantages and drawbacks, in terms of control success, energy costs, and experimental feasibility~\cite{STA_review}. Counterdiabatic driving (CD, also called transitionless quantum driving), introduced in~\cite{JPA_Berry} is one of these methods and involves the introduction of an additional external Hamiltonian such that the adiabatic eigenstates of the original Hamiltonian are the exact solutions of the combined Hamiltonian. The method has attracted much attention~\cite{PRL_delCampo,PRX_Deffner,PRE_Takahashi,PRL_Funo,PRL_Campbell,JPA_Steve,PRA_Takahashi,SciRep_Alan,PRA_Alan,JPA_Alan} and has been implemented experimentally~\cite{PRL_Zhang,NatComm_delCampo,OptLett_Alan}. Recently, incorporating STA techniques to speed-up QHEs has been proposed in different physical systems as working mediums~\cite{PRE_Gong1,EPL_Obinna,PRE_Obinna,PRE_Obinna2,NJP_Steve,arXiv_delCampo,Entropy_Beau,SciAdv_Deng,SciRep_delCampo,PRE_Berakdar,PRA_Cavina1,PRA_Cavina2}. Nevertheless, such protocols come with their own energetic costs that also need to be carefully accounted for when evaluating the performance of a cycle~\cite{EPL_Obinna,PRE_Obinna,PRE_Obinna2,NJP_Steve,arXiv_delCampo,Entropy_Beau,SciAdv_Deng,SciRep_delCampo,PRE_Berakdar,NJP_Tobalina}. 

In this work, we specifically consider a quantum Otto cycle whose working medium is constituted by spin-$1/2$ particles. We use an STA scheme based on CD in order to mimic an adiabatic cycle at a finite-time. We evaluate the performance of a proposed STA engine by properly including the cost of the application of a CD Hamiltonian, and we investigate the trade-offs between these STA costs and the thermodynamic figures of merit. Moreover, we also compare the performance of an STA engine with those of the adiabatic and non-adiabatic quantum Otto cycles, and we show that the improvement provided by the CD is significant from different perspectives. The assessment of the cost of STA driving in a thermodynamical process is an ongoing debate in the literature, and the method to include the STA cost we employ throughout this work is not unique~\cite{PRA_Zhang,PRA_Muga,PRA_Calzetta,arXiv_Guff,EPL_Obinna,PRE_Obinna,NJP_Tobalina}. We also briefly touch on these alternative cost evaluation methods in the present case.

This paper is organized as follows. In Sec.~\ref{prelim} we introduce the concepts that are central to this work such as the CD scheme, details of the quantum Otto cycle and how to characterize its performance with and without the presence of a CD Hamiltonian. Sec~\ref{sec:singlespin} and Sec.~\ref{sec:twospin} presents our main results on the STA quantum Otto engine and its performance, together with their comparison to the adiabatic and non-adiabatic engine cycles for single and two-spin working mediums, respectively. We conclude in Sec.~\ref{sec:conclusion}.
\section{Preliminaries}\label{prelim}
\subsection{Counterdiabatic driving}\label{sec:ca_drive}
Consider a system whose time evolution is governed by a time-dependent Hamiltonian $H_0(t)$. If the rate of change of the Hamiltonian is infinitely slow, i.e. adiabatic, the system will have no transitions between the eigenstates. In the opposite limit, where the Hamiltonian is varied at very short time scales, there will be excitations induced in between the energy levels and thus the system will no longer follow adiabatically the instantaneous eigenstates of the Hamiltonian. In order to mimic the adiabatic evolution at a finite time, it is possible to adopt a CD scheme in which an additional Hamiltonian, $H_{\text{CD}}(t)$, is introduced so that the instantaneous eigenstates of the total Hamiltonian $H(t)\!=\!H_0(t)+H_{\text{CD}}(t)$ follows the adiabatic solution of $H_0(t)$. The form of $H_{\text{CD}}(t)$ can be exactly calculated as~\cite{JPA_Berry}
\begin{equation}\label{berry_cd}
H_{\text{CD}}(t)=i\hbar\sum_n\left(\partial_t|n(t)\rangle\langle n(t)|-\langle n(t)|\partial_tn(t)\rangle |n(t)\rangle\langle n(t)|\right),
\end{equation}
where $|n(t)\rangle$ is the $n^{th}$ eigenstate of the original Hamiltonian $H_0(t)$. The requirement of the knowledge of the instantaneous spectrum of $H_0(t)$ complicates the calculation of the CD Hamiltonian in many-body systems. However, there are different ways to overcome this difficulty with the use of alternative techniques that allows one to determine the driving scheme without the necessity of diagonalizing $H_0(t)$~\cite{PRL_delCampo,PRX_Deffner}.

\subsection{Quantum Otto Cycle}\label{sec:cycle}
The quantum Otto cycle consists of two quantum adiabatic and two quantum isochoric stages \cite{PRE_Quan1,PRE_Quan2}.
While the classical adiabatic process only demands isentropic transformation, the quantum adiabaticity condition is more restrictive.
It requires that the populations of the energy levels to remain unchanged. In simple cases where it is clearly recognized that all the energy
gaps in a system shrinks during quantum adiabatic transformation, one could say this is an effective expansion by associating
the control parameter of the adiabatic transformation to the reciprocal of some effective size of the system. Similar terminology
may be used for naming one quantum adiabatic branch of the cycle as compression. For more complicated cases where
energy gaps change inhomogeneously with the control parameter then one could still name the quantum adiabatic branches
as compression or expansion, effectively by examination of the work output of the system. If the work is produced by the system
during the quantum adiabatic process then it could be named expansion, while if the work is done on the system then it could
be named as compression. Details of the overall cycle are as follows: 
\begin{itemize}
\setlength\itemsep{-0.76em}
\item \underline{\textit{Heating:}} The working substance is put in contact with a hot reservoir at temperature $T_1$. During this time, $\tau_1$, heat is absorbed by the system, $\langle Q_1\rangle$, and it thermalizes to the temperature of the bath. No work is performed in this branch. \newline
\item \underline{\textit{Expansion:}} The system is isolated from the heat bath and undergoes an isentropic expansion process in which the time-dependent parameter in the working fluid Hamiltonian is decreased in time $\tau_2$. During this stage work is extracted from the working substance with an amount of $\langle W_2\rangle$. \newline
\item \underline{\textit{Cooling:}} The working substance is put in contact with a cold reservoir at temperature $T_2$. During this time, $\tau_3$, heat is transferred from the system to the bath, $\langle Q_3\rangle$, and it thermalizes to the temperature of the bath. No work is performed in this branch. \newline
\item \underline{\textit{Compression:}} The system is isolated from the heat bath and undergoes an isentropic compression process in which the time-dependent parameter in the working fluid Hamiltonian is increased in time $\tau_4$. During this stage work is performed on the working substance with an amount of $\langle W_4\rangle$.
\end{itemize}
The average absorbed and removed heat, $\langle Q_1\rangle$ and $\langle Q_2\rangle$, is calculated as $\text{Tr}(\Delta\rho H)$, where $\Delta\rho$ is the change in the density matrix of the working medium between the beginning and end of the heating and cooling stages, respectively. Similarly, the average produced work by and the work done on the working medium, $\langle W_2\rangle$ and $\langle W_4\rangle$, are calculated as $\text{Tr}(\rho \Delta H)$, where $\Delta H$ is the change in the Hamiltonian of the system between the beginning and end of the expansion and compression stages, respectively. During these stages, the evolution of the working fluid is determined by the von-Neumann equation $\dot{\rho}(t)=-i[H(t),\rho(t)]$.

\subsection{Performance of the engine}\label{sec:performance}
Assessment of the performance of an engine is generally quantified by the efficiency which is the ratio of output energy to the input energy and the power of the cycle which is the delivered energy during the cycle duration. Considering an adiabatic cycle, these quantities are given as follows
\begin{align}
\eta_{\text{A}}&=\frac{\langle W_{2}\rangle+\langle W_{4}\rangle}{\langle Q_1\rangle},  &   P_{\text{A}}&=\frac{\langle W_{2}\rangle+\langle W_{4}\rangle}{\tau_{\text{cycle}}}.
\end{align}

However, CD requires the introduction of an engineered additional Hamiltonian which needs to be implemented externally. 
Therefore, performance of the STA engine needs to be determined by also including the cost of applying the CD Hamiltonian during the expansion and compression strokes. In this case, the efficiency and power of the engine are modified as follows~\cite{EPL_Obinna, PRE_Obinna, PRE_Obinna2, NJP_Steve}
\begin{equation}\label{sta_eff}
\eta_{\text{STA}}=\frac{\langle W_2^{\text{STA}} \rangle+\langle W_4^{\text{STA}}\rangle}{\langle Q_1\rangle+\langle \dot{H}_{\text{CD}}^2 \rangle_{\tau}+\langle \dot{H}_{\text{CD}}^4\rangle_{\tau}}
\end{equation}
and
\begin{equation}\label{sta_power}
P_{\text{STA}}=\frac{\langle W_2^{\text{STA}}\rangle+\langle W_4^{\text{STA}}\rangle-\langle \dot{H}_{\text{CD}}^2\rangle_{\tau}-\langle \dot{H}_{\text{CD}}^4\rangle_{\tau}}{\tau_{\text{cycle}}},
\end{equation}
where $\tau_{\text{cycle}}$ is the total cycle time of the engine and we characterize the cost as 
\begin{equation}\label{maincost}
\langle \dot{H}_{\text{CD}}^i\rangle_{\tau}\!=\!\int_0^{\tau}\langle \dot{H}_{\text{CD}}^i(t)\rangle dt
\end{equation}
with $i\!=\!2, 4$, which is the sum of the average of the time derivative of the CD Hamiltonian over the driving time. The motivation behind the definition of our cost function is as follows. Physically, during a non-ideal (finite time) adiabatic transformation emergence of quantum coherences in the energy basis is generally associated with quantum friction, in the sense that these coherences decrease the efficiency of a thermodynamics process. Counterdiabatic driving terms provide a quantum lubricant~\cite{PRE_Feldmann,feldmann_quantum_2006,PRL_Plastina} suppressing the emergence of energy coherences during such finite time transformations. Hence the cost above, is the additional work associated with the introduction of such CD terms against friction, which is required to drive a working system at a finite rate without invoking any energy coherences. A similar approach in a different physical setting has also been presented in~\cite{arXiv_Dann}. Note that, we calculate the expectation value using the states driven by the original Hamiltonian, i.e. by using the solution of $\dot{\rho}_0(t)\!=\!-i[H_0(t), \rho_0(t)]$. We give a detailed reasoning and derivation of this definition in Appendix~\ref{costappendix}. Again, we would like to stress that the way to determine the cost of an STA protocol is still an ongoing debate in the literature and there is no single definition. 

Since the STA scheme enables us to mimic the adiabatic evolution, the total work output in these equations is equal to that of the adiabatic cycle $\langle W_{A}\rangle=\langle W_2^{\text{STA}}\rangle+\langle W_4^{\text{STA}}\rangle$. Therefore, in the absence of the CD Hamiltonian and the adiabatic evolution of the system, $\eta_{\text{STA}}$ reduces to $\eta_\text{A}$. 

Ideally, expansion and compression strokes of the Otto cycle are assumed to be made adiabatically, i.e. $\tau_2, \tau_4\rightarrow \infty$, in order to avoid any transitions between the eigenstates of the working fluid during these stages which can reduce the work output and therefore the efficiency of the engine. Due to these very long, quasi-static processes, such an adiabatic engine has a vanishingly small power output. It is possible to obtain a finite power from a non-adiabatic engine, however, the total work output from this engine will be diminished due to its finite-time character and it can be given as $P_{\text{NA}}=(\langle W_2\rangle+\langle W_4\rangle)/\tau_{\text{cycle}}$. While calculating the power, following the common approach in the literature~\cite{PRE_Quan1,NJP_Rezek,Entropy_Kosloff}, we will assume that the bottleneck for the total cycle time are the times spent in expansion and compression stages of the cycle, i.e. $\tau_{1,3}\!\ll\!\tau_{2,3}$. Therefore, we will assume $\tau_{cycle}\!=\!\tau_2+\tau_4\!=\!2\tau$. Finally, the definitions above are not unique and alternative measures are also present~\cite{PRA_Zhang,arXiv_Guff}.

In the following sections, we will investigate the quantum Otto cycle with working mediums consisting of single and two-spin-$1/2$ systems. Mainly, we will compare the performance of the STA engine with the adiabatic and non-adiabatic engine, in order to see if the price we pay due to the introduction of CD Hamiltonian to make the adiabatic cycle at a finite-time still gives us an advantageous engine as compared to non-adiabatic one~\cite{EPL_Obinna, PRE_Obinna, PRE_Obinna2, NJP_Steve}.

\section{Single-spin working medium}\label{sec:singlespin}
We begin by assuming that our working medium for the quantum Otto cycle is a single spin in an arbitrary magnetic field described by the Hamiltonian
\begin{equation}\label{h0}
H_0(t)=\vec{b}(t)\cdot\vec{\sigma},
\end{equation}
where $\vec{b}(t)$ is the vector characterizing the external time-dependent magnetic field and $\vec{\sigma}\!=\!\{\sigma_x, \sigma_y, \sigma_z\}$ are the usual spin-$1/2$ Pauli matrices. The total work output and the corresponding efficiency if the work strokes are performed adiabatically 
are as follows~\cite{PRL_Kieu,EPJD_Kieu}
\begin{align}
\langle W_{\text{A}}\rangle &=\langle W_2\rangle+\langle W_4\rangle =-(|\vec{b}_i|-|\vec{b}_f|) \left[\tanh\left(\frac{|\vec{b}_i|}{T_1}\right)-\tanh\left(\frac{|\vec{b}_f|}{T_2}\right)\right], \label{a_work} \\
\eta_{\text{A}} &=1-\frac{|\vec{b}_f|}{|\vec{b}_i|}, \label{a_eff}
\end{align}
where the subscript $A$ denotes the adiabatic evolution, $T_1$ ($T_2$) is the temperature of the hot (cold) bath, and $|\vec{b}_i|$ ($|\vec{b}_f|$) is the initial (final) magnitude of the external magnetic field vector at the beginning (end) of the adiabatic branch,  respectively. In order to have a working engine cycle, we need to satisfy the conditions $|\vec{b}_f|\!<\!|\vec{b}_i|$, and more importantly 
\begin{equation}\label{Tconst}
T_1>T_2\frac{|\vec{b}_i|}{|\vec{b}_f|}. 
\end{equation}
Note that the latter constraint is tighter than the classical one where only $T_1\!>\!T_2$ is required~\cite{PRL_Kieu}. However, due to the quasi-static nature of the adiabatic change of the external magnetic field, power delivered by this engine cycle vanishes.

\begin{figure*}
{\bf (a)} \hskip0.65\columnwidth {\bf (b)}\hskip0.65\columnwidth {\bf (c)}\\
\includegraphics[width=0.7\columnwidth]{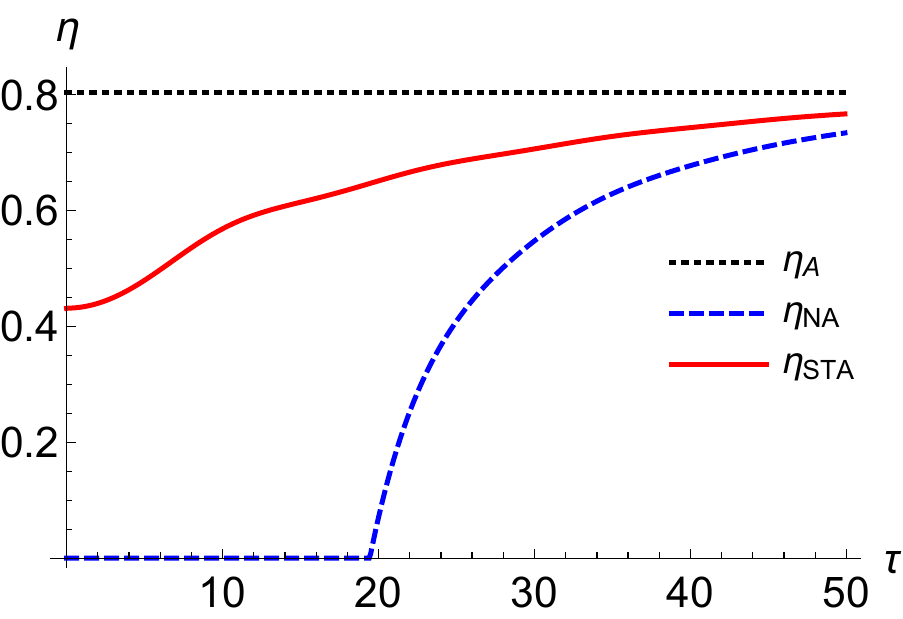}\includegraphics[width=0.7\columnwidth]{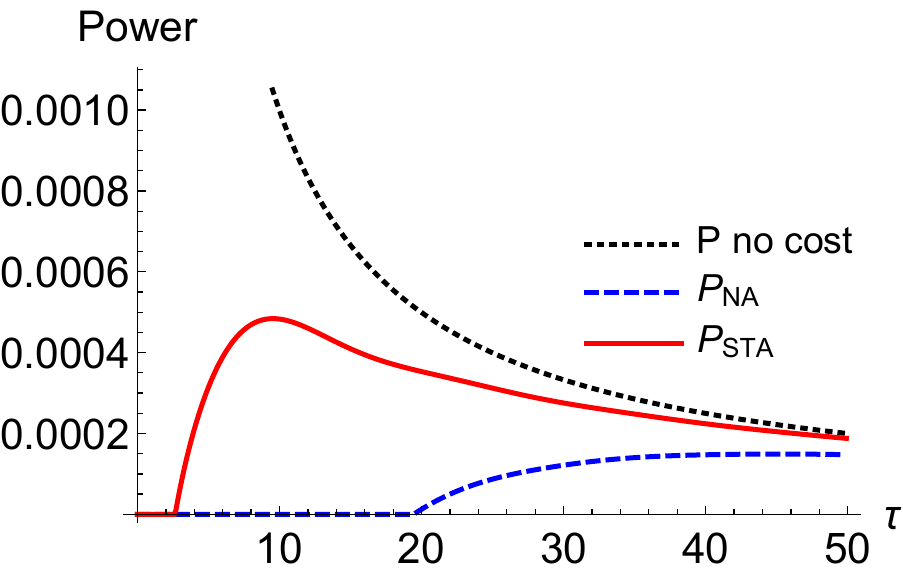}\includegraphics[width=0.7\columnwidth]{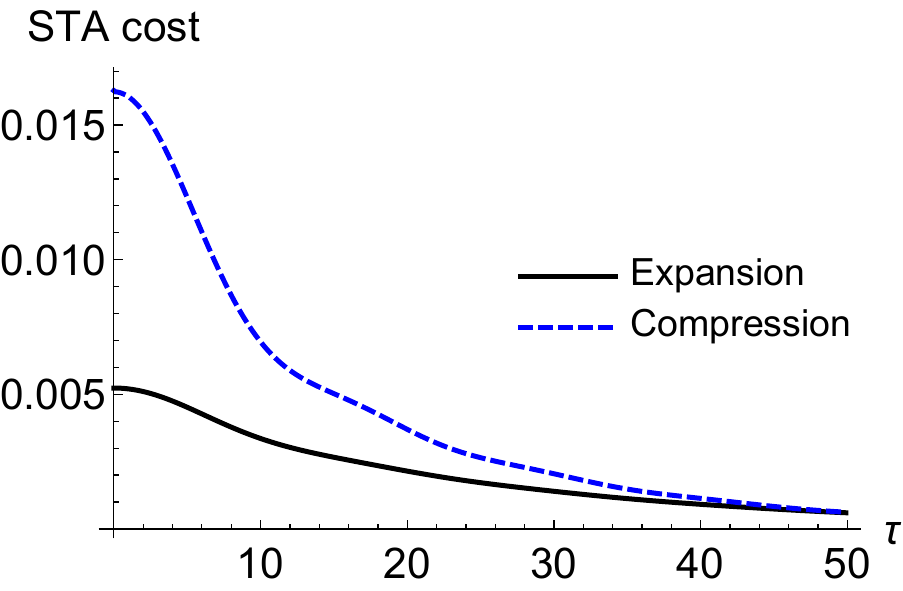}\\
\caption{Efficiency {\bf (a)} and power {\bf (b)} of the single-spin quantum Otto engine vs. the driving time $\tau$ for adiabatic (black dotted), non-adiabatic (blue dashed) and STA (red thick) work strokes. In {\bf (b)} black dotted line denotes the STA power output with no cost for reference. The cost of STA is presented in {\bf (c)} for the expansion (black solid) and compression (blue dashed) strokes. The system parameters are $b_i\!=\!\{0.1, 0, 0.5\}$, $b_f\!=\!\{0.1, 0, 0\}$, $T_1\!=\!10$ and $T_2\!=\!1$.}
\label{singlespin}
\end{figure*}
\begin{figure*}
{\bf (a)} \hskip0.65\columnwidth {\bf (b)}\hskip0.65\columnwidth {\bf (c)}\\
\includegraphics[width=0.7\columnwidth]{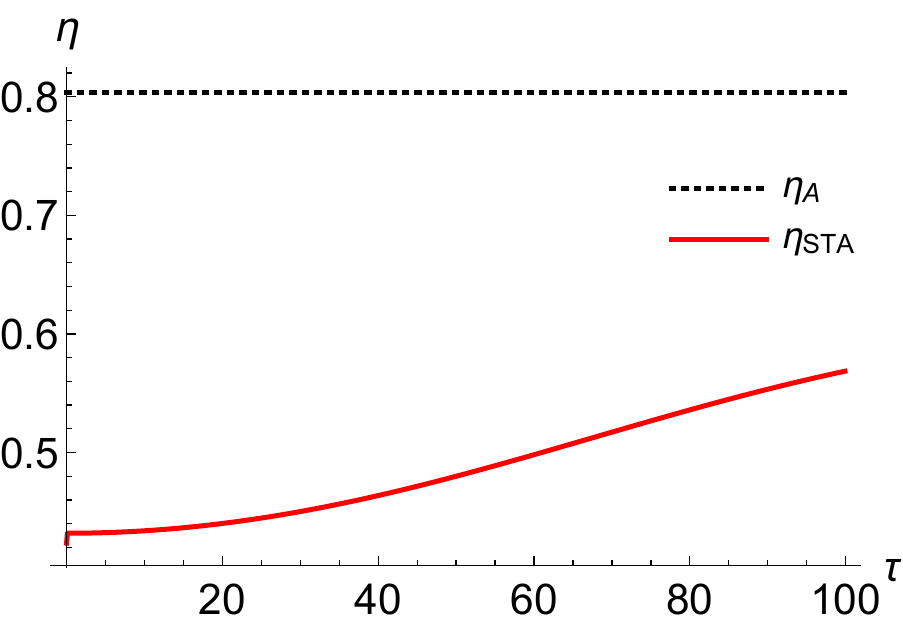}\includegraphics[width=0.7\columnwidth]{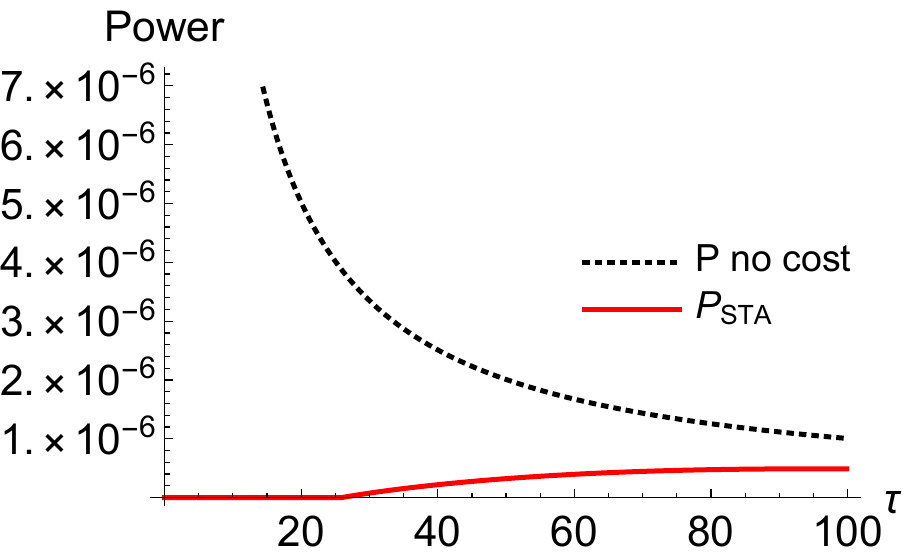}\includegraphics[width=0.7\columnwidth]{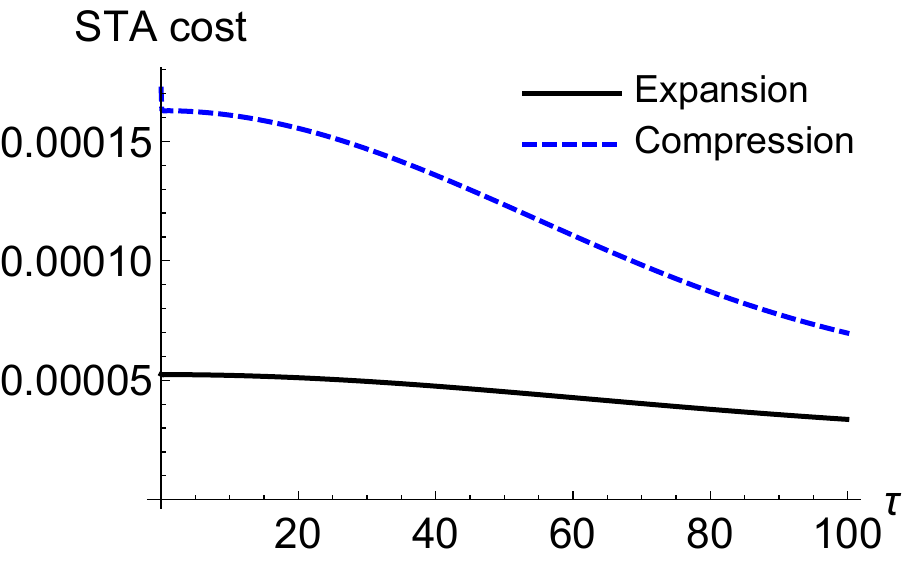}\\
\caption{Efficiency {\bf (a)} and power {\bf (b)} of the single-spin quantum Otto engine vs. the driving time $\tau$ for adiabatic (black dotted) and STA (red thick) work strokes. In {\bf (b)} black dotted line denotes the STA power output with no cost for reference. The cost of STA is presented in {\bf (c)} for the expansion (black solid) and compression (blue dashed) strokes. The system parameters are $b_i\!=\!\{0.01, 0, 0.05\}$, $b_f\!=\!\{0.01, 0, 0\}$, $T_1\!=\!10$ and $T_2\!=\!1$.}
\label{singlespin_smallgap}
\end{figure*}

Application of the introduced STA scheme in Sec.~\ref{sec:ca_drive} to the present system at hand in order to mimic the the adiabatic evolution during the work strokes at a finite time, gives us the following CD Hamiltonian~\cite{JPA_Berry, PRE_Takahashi}.
\begin{equation}\label{CD}
H_{\text{CD}}(t)=\frac{\vec{b}(t)\times\dot{\vec{b}}(t)}{2|\vec{b}(t)|^2}\cdot\vec{\sigma}=\vec{b}_{\text{CD}}(t)\cdot\vec{\sigma}.
\end{equation}
By construction, the driving Hamiltonian $H_{\text{CD}}$ needs to vanish at the beginning and at the end of the driving time, i.e. $H_{\text{CD}}(t\!=\!0, \tau)\!=\!0$, which in turn imposes the condition on the vector $|\vec{b}_{\text{CD}}(t\!=\!0, \tau)|\!=\!|\dot{\vec{b}}(t)|/|\vec{b}(t)|\sin(\theta)\!=\!0$. There are two options to satisfy this condition: i) the angle $\theta=0$ or $\pi$, and ii) $|\dot{\vec{b}}(t=0)|\!=\!|\dot{\vec{b}}(t=\tau)|=0$. The former yields a trivial fixed point condition~\cite{PRE_Takahashi}, which we will not focus on in this work. It is possible to satisfy the latter boundary conditions by assuming the following time-dependence profile of $\dot{\vec{b}}(t)$
\begin{equation}\label{sta_bc_d}
\dot{\vec{b}}_{j}(t)=C_j\frac{t}{\tau^2}\left(1-\frac{t}{\tau}\right),
\end{equation}
which in turn gives the external magnetic field profile as follows 
\begin{equation}\label{sta_bc}
\vec{b}_{j}(t)=D_j+C_j\frac{t^2}{\tau^2}\left(\frac{1}{2}-\frac{t}{3\tau}\right),
\end{equation}
where $j\!=\!{x, y, z}$ are the components of the vector $\dot{\vec{b}}(t)$, and $C_j$ and $D_j$ are arbitrary constants that determines the initial and final values of the components of the external magnetic field in the adiabatic branches. An optional boundary condition to ensure the smoothness of the external field trajectory is $\ddot{\vec{b}}(t\!=\!0, \tau)\!=\!0$~\cite{arXiv_delCampo}. Although we did not aim specifically to satisfy this last constraint in the first place, our simple choice of external field profile, Eq.~(\ref{sta_bc}), automatically satisfies it.

We now focus our attention on a specific single-spin model that enables us to explicitly discuss the figures of merit introduced in the previous section and compare them for the cases of adiabatic, non-adiabatic and STA engines. For this purpose, we will restrict the Hamiltonian presented in Eq.~(\ref{h0}) to the following one, which describes the Landau-Zener (LZ) model
\begin{equation}
H_{\text{LZ}}=b_x\sigma_x+b_z(t)\sigma_z,
\end{equation}
where $b_x$ is the minimal splitting between the energy levels and $b_z(t)$ is the time-dependent external field. We would like to note that, the aforementioned trivial solution to the boundary conditions that the corresponding CD Hamiltonian needs to satisfy, does not exist in the case of the LZ model. Therefore, both $b_x$ and $b_z(t)$ need to have the form presented in Eq.~(\ref{sta_bc}) with only $C_x\!=\!0$ since we explicitly assume $b_x$ is time-independent. The corresponding CD Hamiltonian can be determined as 
\begin{equation}
H_{\text{CD}}^{\text{LZ}}(t)=-\frac{b_x\dot{b_z}(t)}{2(b_x^2+b_z^2(t))}\sigma_y.
\end{equation}
The LZ model has an avoided energy level crossing at the point where $b_z(t)\!=\!0$ and it has been shown that the energy gap of the system is highly relevant to the cost of the STA driving scheme \cite{PRL_Campbell}. Therefore, we discuss the performance of the finite-time STA engine in relation to the magnitude of the minimal energy gap, and we determine the operating limits of it accordingly. Such an analysis has the potential to be extended to many-body critical systems, such as Ising \cite{PRL_delCampoIsing} and Lipkin-Meshkov-Glick~\cite{PRL_SteveLMG} models due to their close connection to the LZ model.

We present our results on the efficiency and power of the finite-time single-spin STA Otto cycle in Figs.~\ref{singlespin} and~\ref{singlespin_smallgap}, together with the corresponding adiabatic and non-adiabatic counterparts for comparison. We also explicitly present the cost of applying the CD Hamiltonian in expansion and compression strokes of the cycle. 

Fig.~\ref{singlespin} displays the results when LZ model parameters are set such that $b_x\!=\!0.1$ and change in $b_z$ is between $0.5$ and $0$ in the expansion/compression strokes with the time-dependence presented in Eq.~(\ref{sta_bc}) with $T_1/T_2\!=\!10$. The adiabatic efficiency for the quantum Otto cycle with these parameters is $\eta_\text{A}\!\approx\! 0.8$. We observe in Fig.~\ref{singlespin} {\bf (a)} that the efficiency of the STA engine has a finite value close to the adiabatic engine at very short driving times while the non-adiabatic engine is unable to generate work output. This efficiency lag until $\tau\!\approx\! 19.5$ in the case of non-adiabatic engine is due to the irreversible entropy production when the working fluid experiences a fast driving during the work strokes~\cite{PRE_Feldmann,PRL_Plastina,PRE_Shiraishi,arXiv_Paterson,EPJD_Selcuk}. The STA technique we employ for our engine, suppresses such unwanted irreversible entropy production at the cost of implementing the external CD Hamiltonian, which is responsible for the deviation of the STA engine efficiency from the adiabatic one at short times. The total cost during the cycle does not have a highly significant impact on the efficiency of the STA engine; it brings down $\eta_{\text{STA}}$ from $\eta_\text{A}$ to a minimum of $\approx 0.43$ for driving times as short as $\tau\!=\!0.001$. This corresponds to a $46\%$ loss in the efficiency, however the cycle time is significantly shortened. Naturally, as the driving time is increased the adiabatic limit is recovered for both STA and non-adiabatic engines, since the evolution of the working fluid becomes quasi-static.

The advantage of the STA engine is more pronounced when we look at the generated power out of the cycle. As mentioned before, the adiabatic engine has a very small power output due to very long expansion/compression stroke times. The ability of the STA engine to generate the adiabatic work output at a finite time significantly enhances the delivered power at finite-times, as shown in Fig.~\ref{singlespin} {\bf (b)}.However, for very short driving times, the STA engine also fails to deliver a power output, due to the high cost of introducing the CD Hamiltonian to the system. As the costs decrease below the total work output with increased driving time, we are able to obtain a finite power. In the case of a non-adiabatic engine, based on the aforementioned reasons for the efficiency lag, the power also displays a lag until the same driving time. Therefore, the proposed STA engine outperforms both the adiabatic and non-adiabatic engines in terms of the output power of the cycle.

\begin{figure*}
{\bf (a)} \hskip0.65\columnwidth {\bf (b)}\hskip0.65\columnwidth {\bf (c)}\\
\includegraphics[width=0.7\columnwidth]{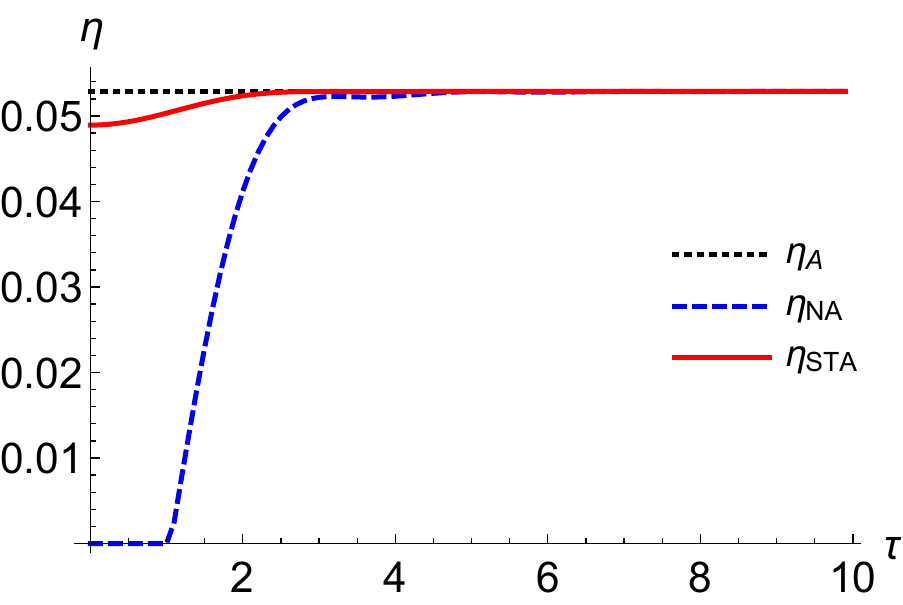}\includegraphics[width=0.7\columnwidth]{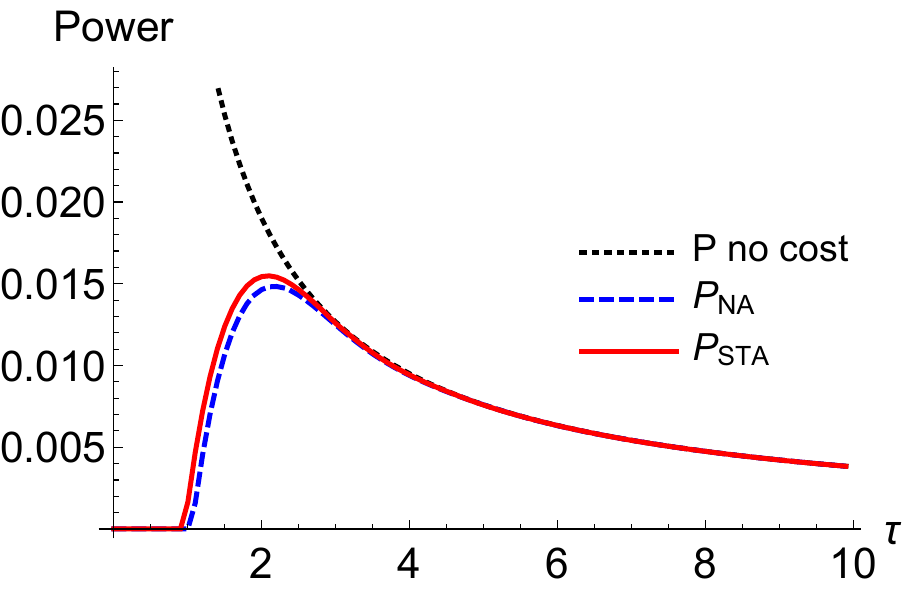}\includegraphics[width=0.7\columnwidth]{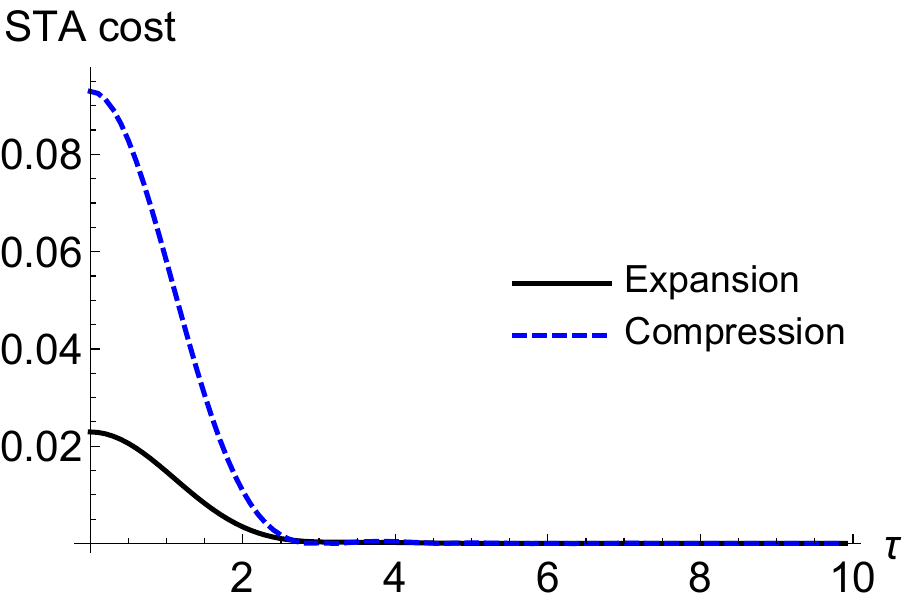}\\
\caption{Efficiency {\bf (a)} and power {\bf (b)} of the two-spin quantum Otto engine vs. the driving time $\tau$ for adiabatic (black dotted), non-adiabatic (blue dashed) and STA (red thick) work strokes. In {\bf (b)} black dotted line denotes the STA power output with no cost for reference. The cost of STA is presented in {\bf (c)} for the expansion (black solid) and compression (blue dashed) strokes. The system parameters are $\gamma=0.7$, $h_i\!=\!0.5$, $h_f\!=\!0$, $T_1\!=\!10$ and $T_2\!=\!1$.}
\label{twospin}
\end{figure*}

We now turn our attention to a more detailed analysis of the STA cost in the expansion and compression stages of the cycle. From Fig.~\ref{singlespin} {\bf (c)}, we observe that the cost during the compression process is higher than that of the expansion process. Recall that the former process involves increasing the energy gap of the spin while in the latter, we do the opposite. From these results we can infer that the suppression of non-equilibrium entropy production in the system is harder to sustain when the energy gap is opening in contrast to when it is closing. It may be possible to intuitively understand this result as follows: when the gap is at its minimum, any small perturbation in the system can easily induce transitions between the levels. The compression stroke starts right at this minimum, therefore suppressing the transitions near it can cost more than when the energy levels start far apart and are brought together where small perturbations have less of an impact. It may also be argued that the time-dependence profile of the external magnetic field also affects these costs. All in all, we think that the reason behind this behavior calls for a more detailed physical understanding.

Before moving on to the LZ model engine with a smaller gap, we would like to clarify one point. To have the benchmark adiabatic engine with the same efficiency as in the previous results but with a smaller gap, we need to sweep the magnetic field in the $z$-direction in a more restricted range. The reason behind this is as follows: the minimum gap is determined by $b_x$, and the magnitude of the initial and final magnetic field vector is determined by the range of $b_z$, after $b_x$ is fixed. Since the efficiency of the adiabatic cycle explicitly depends on the ratio, $|\vec{b}_i|/|\vec{b}_f|$, when $b_x$ is decreased, the values within which $b_z$ is varied also needs to get smaller in order to maintain the same efficiency. Moreover, keeping the range of $b_z$ when $b_x$ is decreased not only changes the efficiency of the engine, but it also modifies the ratio of the temperatures of the hot and cold baths so that the cycle can operate as an engine due to the constraint given in Eq.~(\ref{Tconst}). Therefore, to be able to make a reasonable comparison with the previous case with a larger gap, we introduce the restriction on the $b_z$ range.

Fig.~\ref{singlespin_smallgap} presents our results on the same single-spin engine but with a smaller minimum energy gap, in order to show the effects of the gap magnitude between the energy levels. We set the model parameters as $b_x\!=\!0.01$, and the change in $b_z$ is between $0.05$ and $0$ with $T_1/T_2\!=\!10$ again. To begin with, it is not possible to have an operating non-adiabatic engine for this case with the considered driving times. Since the energy gap is very small, even the smallest deviation from quasi-static behavior very easily causes unwanted transitions between the energy levels which makes it impossible for a non-adiabatic engine to generate work output even when we consider driving times that are twice as long (cf. y-axes of Fig.~\ref{singlespin} and Fig.~\ref{singlespin_smallgap}). We have confirmed that a non-adiabatic engine can only begin to operate for driving times around $\tau\!\approx\! 200$. 

The STA engine, on the other hand, works quite well with $\eta_{\text{STA}}\approx 0.43$ for driving times as short as $\tau\!=\!0.001$, similar to the previous case with a larger energy gap. Realizing that the work output is lowered due to smaller $|\vec{b}_i|$ and $|\vec{b}_f|$, this is only possible if the cost related to the STA is also lowered in this case, as can be seen from Fig.~\ref{singlespin_smallgap} {\bf (c)}. This may seem contrary to the general and reasonable intuition that when the gap is smaller it is more costly to prevent the excitations from occurring at a finite-time drive~\cite{PRL_Campbell}. However, we think that the reason behind the lowered cost is the fact that we sweep a more restricted range of external magnetic field. The struggle of the CD scheme with the smaller gap can still be seen in the slower convergence of the efficiency and cost functions to their adiabatic values. In other words, even though we do not see an increase in the magnitude of the STA cost when the energy gap is lowered, its slower decaying behavior as a function of driving time is a sign of a harder to manage driving process. The reduction in the power of this cycle for the STA engine can again be explained by the decreased work output resulting from the choice of smaller external magnetic field vector lengths.

We have stated that the characterization of the cost of CD driving that we have adopted in the previous section is not unique, and there are other ways to asses the performance of the engine. We would like to briefly comment on these different ways of quantifying the cost and how they relate to our results. To begin with, we may compare our cost definition given in Eq. (\ref{maincost}) with the one introduced in \cite{EPL_Obinna}. The latter can be obtained by replacing the time derivative of the CD Hamiltonian inside the integral, $\langle \dot{H}_{\text{CD}}(t)\rangle$ in Eq. (\ref{maincost}) with $\langle H_{\text{CD}}(t)\rangle/\tau$. When we made such a definition change and apply it in our model, we have seen that while the qualitative behavior remains the same, there is a significant quantitative decrease in the cost. However, we also would like to point out the subtle difference between the states we use to calculate the expectation value, which are driven by $H_0(t)$, and those that are used in \cite{EPL_Obinna}, which are the eigenstates of an unitary equivalent of $H_{\text{CD}}(t)$. Furthermore, analyzing the cost definition in \cite{PRA_Zhang} based on the Frobenius norm of the driving Hamiltonian for the present case, we have observed that the cost values exceed the ones presented in this work. On the other hand, inclusion of the cost in the efficiency and power of the engine cycle is also a topic of debate where alternatives are present. One proposal in this direction is to include the cost needed to implement the CD Hamiltonian to the numerator of the efficiency~\cite{arXiv_Guff}, as opposed to subtracting it from the denominator as we did in Eq.~(\ref{sta_eff}). We would like to stress that even in the case of such a definition change, the advantages of the STA engine still persist.  

\section{Two-spin engine}\label{sec:twospin}
A natural direction to follow at this point is to consider a working medium that is composed of a larger number of particles. Quantum Otto engines that are made out of two spins are extensively investigated in the literature~\cite{PRE_Ferdi, PRE_Ferdi2, EPJP_Selcuk, PRA_Gabriele}, however these works are restricted to the assumption of a quasi-static (adiabatic) cycle. In what follows, we will consider a two-spin STA engine by applying the CD scheme introduced in~\cite{PRE_Takahashi}. We assume that the self-Hamiltonian of the two-qubit working medium made out of qubits $a$ and $b$, is described by the $XY$ Hamiltonian in a transverse magnetic field,
\begin{equation}
H_0(t)=J_x(t)\sigma_x^a\sigma_x^b+J_y(t)\sigma_y^a\sigma_y^b+h(t)(\sigma_z^a+\sigma_z^b),
\end{equation}
where $J_x(t)$ and $J_y(t)$ are coupling strengths in $x$ and $y$ directions, respectively, and $h(t)$ is the external magnetic field strength. To perform the STA for such a working substance in the expansion and compression branches, the CD Hamiltonian takes the following form~\cite{PRE_Takahashi}
\begin{equation}\label{CD_two_general}
H_{\text{CD}}(t)=\frac{1}{2}\frac{h(\dot{J_x}-\dot{J_y})-\dot{h}(J_x-J_y)}{4h^2+(J_x-J_y)^2}(\sigma_x^a\sigma_y^b+\sigma_y^a\sigma_x^b).
\end{equation}

It is possible to consider two different scenarios for the expansion and compression stages in this model, which can be specified as follows: \textit{(i)} varying the magnetic field while keeping the interaction strength constant, and \textit{(ii)} fixing the magnetic field while varying the interaction strength between the two spins. However, we will not consider the latter case in the present work and investigate the engine cycle where the work strokes are performed by a time-dependent external magnetic field to be able to make a self-contained analysis together with the previous section. The corresponding analytical expressions for the work output and the efficiency for the adiabatic version of such an engine are presented in Appendix~\ref{twospinappendix}.

In the former case, \textit{(i)}, to avoid the trivial fixed point condition in the CD Hamiltonian, we need to introduce an anisotropy between the $x$ and $y$ directions, which we realize by setting $J_x(t)\!=\!J_x=1+\gamma$ and $J_y(t)\!=\!J_y=1-\gamma$. Note that this model is merely the anisotropic $XY$ model in a transverse magnetic field. Plugging these interaction parameters between the system qubits in Eq.~(\ref{CD_two_general}), yields the following CD Hamiltonian
\begin{equation}\label{CD_two}
H_{\text{CD}}^{\text{h}}(t)=-\frac{\dot{h}\gamma}{4(h^2+\gamma^2)}(\sigma_x^a\sigma_y^b+\sigma_y^a\sigma_x^b),
\end{equation}
where $\gamma$ is the anisotropy parameter. The STA boundary conditions $H_{\text{CD}}^{\text{h}}(t\!=\!0, \tau)\!=\!0$ implies $\dot{h}(t\!=\!0, \tau)\!=\!0$ which can be realized by the same magnetic field profile that we have considered in Eqs.~(\ref{sta_bc_d}) and (\ref{sta_bc}) in the previous section. Explicitly, we have
\begin{align}
\dot{h}(t) &=E\frac{t}{\tau^2}\left(1-\frac{t}{\tau}\right), \\
h(t) &=F+E\frac{t^2}{\tau^2}\left(\frac{1}{2}-\frac{t}{3\tau}\right),
\end{align}
where $E$ and $F$ are constants that determine the initial and final values of the magnetic field.

We present our results on the efficiency and power of the finite-time two-spin STA Otto engine in Fig.s~\ref{twospin}{\bf (a)} and \ref{twospin} {\bf (b)}, together with the corresponding adiabatic and non-adiabatic counterparts for comparison. The working medium parameters are set so that the externally controlled parameters are as similar as possible to the single-spin case and are as follows $\gamma\!=\!0.7$, the magnetic field, $h(t)$, is varied between $0.5$ and $0$ with $T_1/T_2\!=\!10$. The efficiency lag for the non-adiabatic engine is again present for the short driving times whereas the STA engine attains an efficiency very close to the adiabatic one. As compared to the single-spin engine, the deviation of $\eta_{\text{STA}}$ from $\eta_\text{A}$ is very small, $\approx 7\%$, although the efficiency of the cycle is significantly reduced. On the other hand, the power of the two-spin engine is two orders of magnitude higher,  but the impact of the cost is more pronounced in contrast to the single-spin engine such that the difference between the power outputs of STA and non-adiabatic engines is marginal. The expected convergence of the non-adiabatic and STA engine to the adiabatic values happens at very short times. In fact, the convergence is so quick that, if operating times shorter that $\tau_{\text{cycle}}\!\approx\! 3$, i.e. expansion/compression stroke times around $\tau\!\approx\! 1.5$, are not aimed, one may consider working with a non-adiabatic engine without dealing with the possible complications of the STA scheme.

Further, we present the cost of applying the CD Hamiltonian in Fig.~\ref{twospin} {\bf (c)} and observe that the cost for driving two-spins, with the defined parameters, is an order of magnitude higher than the single-spin driving. Also, the same behavior that was presented and discussed in Fig.s~\ref{singlespin} and \ref{singlespin_smallgap} {\bf (c)} is still present, that is, the STA cost for the compression stroke is higher than that of the expansion cost, even in a more pronounced manner.

\section{Conclusion}\label{sec:conclusion}

We have proposed single and two-spin quantum Otto engines that utilize a proper STA scheme based on CD to achieve an adiabatic work output at finite time. We have characterized the thermodynamic efficiency and power associated with these engines by fully accounting for the cost of external CD Hamiltonian control on systems. Comparing the the same figures of merit calculated for the adiabatic and non-adiabatic engines for the same working medium with the STA engine, we have shown that the STA engine is advantageous in many aspects, by considering specific single and two-spin models, i.e. LZ and anisotropic $XY$ in transverse magnetic field models, respectively. While they show an immediate increase in the power as compared to the adiabatic cycles at the price of a reasonable cost in efficiency, their performance is superior to their non-adiabatic counterparts in both efficiency and power. Interestingly, we have observed that the cost of applying the STA scheme is higher in the compression stroke than in the expansion stroke for all considered cases. We also compared the method we have adopted to characterize the cost with the previously introduced proposals on this subject.

We think that this work contributes well to the recently developing efforts on introduction and characterization of STA quantum Otto engines~\cite{EPL_Obinna, PRE_Obinna, PRE_Obinna2, NJP_Steve}. Moreover, the particular choice of LZ model in the single-spin case offers a promising direction towards systems like the Ising and Lipkin-Meshkov-Glick models. Such critical many-body systems improve the workings of QHEs when considered as the working medium even when no STA protocol is introduced to hinder irreversible entropy production~\cite{NatComm_Campisi}. Finally, very recently a non-adiabatic single-spin QHE is realized in an NMR setup~\cite{arXiv_Paterson,de_assis_quantum_2018}, which may also be a test-bed for the STA engines proposed in this work.

\acknowledgements
B. \c{C}. would like to thank Steve Campbell for many fruitful discussions and insightful comments. The authors would like to acknowledge an enlightening correspondence with Gonzalo Muga and an important comment from Ferdi Alt\i nta\c{s}.

\bibliography{transitionless_qhe}

\onecolumngrid

\appendix

\section{Energy cost of counterdiabatic drive}\label{costappendix}

Here, we will introduce the energy cost calculation for the CD drive in general which is applied to our specific model in the main text. Let us consider a general Hamiltonian with explicit time dependence $H_0(t) := H_0(b(t))$ due to finite time variations of its parameters denoted by $b(t)$. Finite time driving of the system from a given initial state to a target state causes transitions between energy eigenstates. This leads to the building of quantum coherences in the energy basis. Associated additional energy stored in the coherences of the system dissipates in the subsequent thermal stages of engine cycles and hence the term quantum internal friction is coined with the effect~\cite{PRL_Plastina}.  The method of CD driving consists of introducing an additional control term $H_{\text{CD}}(t)$ to the model system such that coherences are not produced during the evolution, and the system remains diagonal in its energy basis. Despite the fact that the evolution becomes free of quantum friction, it is expected that there is an energy cost for such a quantum lubrication~\cite{feldmann_quantum_2006,PRA_Zhang,PRA_Muga,PRA_Calzetta,arXiv_Guff,EPL_Obinna,PRE_Obinna,NJP_Tobalina}. To estimate operational efficiency of quantum heat engines, therefore, one must carefully account for the energy cost of using a CD drive source. The general model of CD driving can be written as~\cite{JPA_Berry}
\begin{eqnarray}
H(t)=H_0(t)+H_{\text{CD}}(t),
\end{eqnarray}
where
\begin{eqnarray}
H_0(t)&=&\sum_nE_n(t)|n(t)\rangle\langle n(t)|,\\
H_{\text{CD}}(t)&=&i\hbar\sum_n\left(\partial_t|n(t)\rangle\langle n(t)|-\langle n(t)|\partial_tn(t)\rangle |n(t)\rangle\langle n(t)|\right),
\label{eq:Hcd}
\end{eqnarray}
where $|n(t)\rangle$ is the $n^{th}$ eigenstate of the Hamiltonian $H_0(t)$. The CD term operates from $t=0$ to $t=\tau$ and 
the system is transferred from an initial state
\begin{eqnarray}\label{eq:init}
\rho(0)=\sum_n p_n(0)|n(0)\rangle\langle n(0)|,
\end{eqnarray}
to a final state $\rho(\tau)$. 

The rate of change of internal energy $J(t)=dU(t)/dt$ with $U(t)=\text{Tr}(\rho(t)H(t))$ of the system during this time evolution is determined by
\begin{equation}
J(t)=\text{Tr}\left(\frac{d\rho(t)}{dt}H(t)\right)+\text{Tr}\left(\rho(t)\frac{dH(t))}{dt}\right).
\end{equation}
Substituting $\dot{\rho}(t)=-i[H(t),\rho(t)]/\hbar$ and separating the terms involving $H_0(t)$ and $H_{\text{CD}}(t)$ drive we write $J(t)=J_0(t)+J_{\text{CD}}(t)$, where
\begin{align}
J_0(t)&=\text{Tr}\left(\rho(t) \frac{dH_0(t)}{dt}\right), & J_{\text{CD}}(t)&= \text{Tr}\left(\rho(t)\frac{dH_{\text{CD}}(t)}{dt}\right).
\end{align}
To obtain the expressions above, we have used the identities $\text{Tr}\left(-i[H(t),\rho(t)]H(t)\right)/\hbar=\text{Tr}\left(-i[H(t),H(t)]\rho(t)\right)/\hbar=0$. 

The time evolution of an initial eigenket $|n(0)\rangle$ of $H_0$ under $H(t)$, becomes~\cite{JPA_Berry}
\begin{eqnarray}
|\psi_n(t)\rangle=U(t,0)|n(0)\rangle=e^{i(\theta(t)+\gamma(t))}|n(t)\rangle,
\end{eqnarray}
where $U(t,0)$ is the propagator for $H(t)$. The dynamical and geometric phases are denoted by $\theta(t)$ and $\gamma(t)$, respectively.
Accordingly, $\rho(t)$ for the initial state in Eq.~(\ref{eq:init}) becomes
\begin{eqnarray}
\rho(t)=\sum_n p_n(0)|n(t)\rangle\langle n(t)|.
\end{eqnarray}
Hence, at all times, the system follows the adiabatic energy eigenstates of $H_0(t)$ without any transitions between them. For such diagonal state in the energy basis of $H_0(t)$ it is possible to show that $J_{\text{CD}}=0$. From the definition in Eq.~(\ref{eq:Hcd}),
we calculate
\begin{eqnarray}
\frac{dH_{\text{CD}}(t)}{dt}=i\hbar\sum_n\left[
|\ddot{n}\rangle\langle n|+|\dot n\rangle\langle \dot n|
-(\langle \dot n|\dot n\rangle+\langle n|\ddot n\rangle)|n\rangle\langle n|-\langle n|\dot n\rangle(|\dot n\rangle\langle n|+|n\rangle\langle\dot n|)
\right],
\end{eqnarray}
where $|\dot n\rangle\equiv|\partial n(t)/\partial t\rangle$ and $|\ddot n\rangle\equiv|\partial^2 n(t)/\partial t^2\rangle$. We suppress the explicit time dependence of $|n(t)\rangle$ for brevity of notation without ambiguity. The diagonal elements are
found to be
\begin{eqnarray}
\left\langle n\left|\frac{dH_{\text{CD}}(t)}{dt}\right|n\right\rangle=\sum_m\langle n|\dot m\rangle\langle \dot m|n\rangle-\langle\dot n|\dot n\rangle,
\end{eqnarray}
where we have used $\partial\langle m|m\rangle/\partial t=\langle \dot m| m\rangle+\langle m|\dot m\rangle=0$. Similarly using $\langle m|n\rangle=\delta_{mn}$ so that $\langle\dot m|n\rangle=-\langle m|\dot n\rangle$ yields 
\begin{eqnarray}
\left\langle n\left|\frac{dH_{\text{CD}}(t)}{dt}\right|n\right\rangle=\sum_m\langle \dot n| m\rangle\langle m| \dot n\rangle-\langle\dot n|\dot n\rangle=0,
\end{eqnarray}
due to the completeness relation $\sum_m|m\rangle\langle m|=\mathds{1}$. On the other hand, we determine $J_0$ for $\rho(t)$ to be
\begin{eqnarray}
J_0(t)=\sum_n p_n(0)\dot E_n(t)+p_n(0)E_n(t)(\langle n|\dot n\rangle+\langle \dot n|n\rangle=\sum_n p_n(0)\dot E_n(t).
\end{eqnarray}
Hence the net energy change of the system under $H(t)$ in $\tau$ is the same with $H_0$ in the adiabatic limit. 

The lack of friction in the energy
transfer into the system under $H(t)$ can be further
argued by considering $[H_0(t),H_{\text{CD}}(t)]$, which is the fundamental cause of quantum internal friction. We can first rewrite $H_{\text{CD}}(t)$ in the form
\begin{eqnarray}
H_{\text{CD}}(t)=i\hbar\sum_{m\neq n}\sum_n |m\rangle\langle m|\dot n\rangle\langle n|,
\end{eqnarray}
which gives
\begin{eqnarray}
[H_0(t),H_{\text{CD}}(t)]=i\hbar\sum_{m\neq n}\sum_n(E_m-E_n)\langle m|\dot n\rangle|m\rangle\langle n|.
\end{eqnarray}
Eventhough $H_0(t)$ and $H_{\text{CD}}(t)$ are not compatible, their commutator becomes zero for $\rho(t)$ which is diagonal in the energybasis of $H_0(t)$. 

The cost of this quantum lubrication can be determined by writing $\rho(t)\!=\!\rho_0(t)-\delta\rho(t)$ and considering the work done by the source of $H_{\text{CD}}(t)$ to compansate the coherences produced in $\rho_0(t)$ in the finite time evolution in $\tau$, where $\rho_0(t)$ is determined by $\dot\rho_0(t)\!=\!-i[H_0(t),\rho_0(t)]$.
We can write 
\begin{eqnarray}
J_{\text{CD}}(t)= \text{Tr}\left(\rho_0(t)\frac{dH_{\text{CD}}(t)}{dt}\right)-\text{Tr}\left(\delta\rho(t)\frac{dH_{\text{CD}}(t)}{dt}\right)=0,
\end{eqnarray}
and identify the instantaneous cost function for the source of $H_{\text{CD}}(t)$ by using $\rho_0(t)$ such that 
\begin{eqnarray}
\langle \dot{H}_{\text{CD}}(t)\rangle:=\text{Tr}\left(\rho_0(t)\frac{dH_{\text{CD}}(t)}{dt}\right),
\end{eqnarray}
which can be integrated to find the total cost during the driving time of the system as
\begin{equation}
\langle \dot{H}_{\text{CD}}(t)\rangle_{\tau}=\int_0^{\tau}\langle \dot{H}_{\text{CD}}(t)\rangle dt.
\end{equation}

\section{Work and efficiency for two-spin engine}
\label{twospinappendix}
Analytical expressions for work and efficiency of the two-spin engine when the external magnetic field is varied and interaction strength between the spins is kept constant are given as follows
\begin{equation}
\langle W_{A}\rangle=\frac{2\left(f_f-f_i\right) M}{\left[\cosh \left(\frac{2 f_f}{T_2}\right)+\cosh \left(\frac{2}{T_2}\right)\right) \left(\cosh \left(\frac{2 f_i}{T_1}\right)+\cosh\left(\frac{2}{T_1}\right)\right]}
\end{equation}
and
\begin{equation}
\langle \eta_{A}\rangle=\frac{-\left(f_f-f_i\right) M}{\left[\sinh \left(\frac{2}{T_1}\right)
   \cosh \left(\frac{2 f_f}{T_2}\right)-\sinh \left(\frac{2}{T_2}\right) \cosh \left(\frac{2 f_i}{T_1}\right)+\sinh \left(\frac{2(T_2-T_1)}{T_1T_2}\right)+f_i M\right]},
\end{equation}
where $f_i=\sqrt{h_i^2+\gamma^2}$, $f_f=\sqrt{h_f^2+\gamma^2}$ and $M$ is defined as 
\begin{equation}
M=\sinh \left(\frac{2 f_i}{T_1}\right) \left[\cosh \left(\frac{2 f_f}{T_2}\right)+\cosh \left(\frac{2}{T_2}\right)\right]-\sinh \left(\frac{2 f_f}{T_2}\right) \left[\cosh \left(\frac{2f_i}{T_1}\right)+\cosh \left(\frac{2}{T_1}\right)\right].
\end{equation}

\end{document}